\def \SAIT #1 #2 {{\em Mem.\ Soc.\ Astron.\ It.\/} {\bf #1}, #2}
\def \MESS #1 #2 {{\em The Messenger\/} {\bf #1}, #2}
\def \ASTRNACH #1 #2 {{\em Astron. Nach.\/} {\bf #1}, #2}
\def \AAP #1 #2 {{\em Astron. Astrophys.\/} {\bf #1}, #2}
\def \AAL #1 #2 {{\em Astron. Astrophys. Lett.\/} {\bf #1}, L#2}
\def \AAR #1 #2 {{\em Astron. Astrophys. Rev.\/} {\bf #1}, #2}
\def \AAS #1 #2 {{\em Astron. Astrophys. Suppl. Ser.\/} {\bf #1}, #2}
\def \AJ #1 #2 {{\em Astron. J.\/} {\bf #1}, #2}
\def \ANNREV #1 #2 {{\em Ann. Rev. Astron. Astrophys.\/} {\bf #1}, #2}
\def \APJ #1 #2 {{\em Astrophys. J.\/} {\bf #1}, #2}
\def \APJL #1 #2 {{\em Astrophys.. J. Lett.\/} {\bf #1}, L#2}
\def \APJS #1 #2 {{\em Astrophys. J. Suppl.\/} {\bf #1}, #2}
\def \APSS #1 #2 {{\em Astrophys. Space Sci.\/} {\bf #1}, #2}
\def \ASR #1 #2 {{\em Adv. Space Res.\/} {\bf #1}, #2}
\def \BAIC #1 #2 {{\em Bull. Astron. Inst. Czechosl.\/} {\bf #1}, #2}
\def \JSQRT #1 #2 {{\em J. Quant. Spectrosc. Radiat. Transfer\/} {\bf #1}, #2}
\def \MN #1 #2 {{\em Mon. Not. R. Astr. Soc.\/} {\bf #1}, #2}
\def \MEM #1 #2 {{\em Mem. R. Astr. Soc.\/} {\bf #1}, #2}
\def \PLR #1 #2 {{\em Phys. Lett. Rev.\/} {\bf #1}, #2}
\def \PASJ #1 #2 {{\em Publ. Astron. Soc. Japan\/} {\bf #1}, #2}
\def \PASP #1 #2 {{\em Publ. Astr. Soc. Pacific\/} {\bf #1}, #2}
\def \NAT #1 #2 {{\em Nature\/} {\bf #1}, #2}
\documentstyle{memsait}
\input epsf.sty
\begin{opening}
\title{GAW (Gamma Air Watch): a novel imaging Cherenkov telescope}
\author{G. Cusumano, G. Agnetta, B. Biondo, O. Catalano, 
S. Giarrusso, G.~Gugliotta, L. La Fata, M.C. Maccarone, 
A. Mangano, T. Mineo, F. Russo, B. Sacco}
\institute{IFCAI Consiglio Nazionale delle Ricerche, Palermo, Italy}
\date{} 
\end{opening}

\begin{document}

\oddpagefooter{}{}{} 
\evenpagefooter{}{}{} 
\ 
\bigskip

\begin{abstract}
GAW (Gamma Air Watch) is a new imaging Cherenkov telescope designed for 
observation of very high-energy gamma-ray sources. 
GAW will be equipped with a 3 meter diameter 
Fresnel lens as light collector and  with an array of 300 
multi-anode photomultipliers at the focal plane. The pixel size will be 
4 arcmin wide for a total field of view of 10.5 degrees. Whith respect to the 
planned imaging Cherenkov telescopes (CANGAROO III, HESS, MAGIC, VERITAS) 
GAW follows a different approach for what concerns both the optical system and
the detection working mode: the Cherenkov light collector is a single 
acrylic flat Fresnel lens (instead of mirrors) that allows to achieve wide 
field of view; the photomultipliers operate in single 
photoelectron counting mode (instead of charge integration).
The single photoelectron counting mode allows to reach a low energy threshold 
of $\sim$ 200 GeV, in spite of the relatively small dimension of the GAW optic 
system.
 
\end{abstract}

\section{Introduction}

At energies around 0.03 TeV the emission from galactic and extragalactic 
sources is too weak to be detected by instruments onboard satellites 
because of their poor effective area. Actually, since the flux of an 
astronomical source decreases as the energy increases, observations at higher 
energy need a huge effective area that is not feasible
with space detectors. 
Only ground-based experiments achieve enough effective area to observe the 
very low intensity and the soft spectra emitted in this extreme energy band. 
Observations can be performed either by detecting the shower of secondary particles 
produced by the interaction of gamma-ray entering into the high atmosphere, 
or by detecting the Cherenkov light emitted by the relativistic charged 
component of the shower. 
The spread of the secondary particles and  
the intrinsic Cherenkov light cone aperture (1.3 degrees in air) allow 
a strong increase of the effective area being the detectors sensible to gamma rays 
whose trajectory is hundreds of metres far from them.
Imaging Cherenkov telescopes, thanks to their large 
collection area ($\sim 10^5$ m$^2$) and to their very high efficiency in 
rejecting the cosmic ray background, have turned out to be the most sensitive 
instruments for the observation of astrophysical sources above 
250 GeV. 

In this paper we present a novel imaging Cherenkov telescope, GAW 
(Gamma Air Watch) designed to observe gamma-ray sources above 
200 GeV. 
The main components of the telescope (optics, focal surface detector 
and operative mode) are described in Sect.2; the performance is presented 
in Sect.3.

\section{GAW: technical description}

Current and next planned imaging Cherenkov telescopes use large  spherical 
mirror reflectors to collect light. 
The field of view (FOV) of these telescopes ranges from 3 to 5 degrees. A 
larger FOV can not be achieved with reflective systems because of the rapid increase of 
aberrations for off-axis angle.  
Since the FOV has the same order of magnitude as the Cherenkov 
light cone, the conventional Cherenkov telescopes can observe only one source 
each time and they must point to a different sky position 
to measure the diffuse background in the FOV loosing on-source observing time. 
These telescopes are not suitable for surveys 
and for discovery of rapid, transient 
phenomena or serendipity TeV sources. A sensitive survey of the galactic plane 
at TeV energies has to be undertaken yet. 
So, the development of wide angle 
optics for Cherenkov telescopes would be very important in the 
very high-energy astronomy. 
Wide FOV observations with a $\sim$200 GeV low energy threshold 
is the goal of the GAW project. 

\begin{figure}
\epsfysize=9cm 
\hspace{3.5cm}\epsfbox{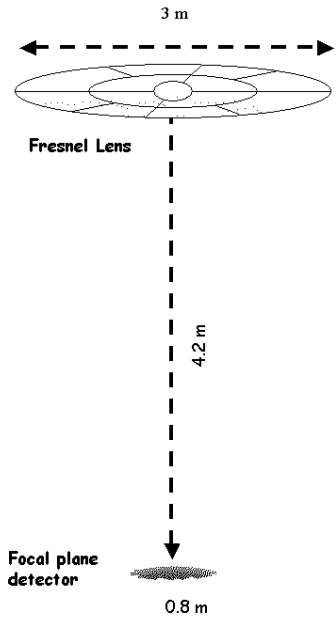} 
\caption[h]{Schematic view of GAW telescope.}
\end{figure}

The main differences
of GAW with respect to the conventional imaging Cherenkov telescopes are: 

\begin{itemize}
\item
optic system: the Cherenkov light collector is a single
acrylic flat Fresnel lens instead of mirrors.
 
\item
photomultiplier operation mode: the photomultipliers work in single
photoelectron counting mode instead of charge integration.
\end{itemize}

A schematic view of the telescope is shown in Fig. 1; the baseline is 
summarized in Table 1.

The Cherenkov light collector of GAW is a Fresnel lens. 
Optic systems with Fresnel lens have the advantage to provide 
large FOV with moderate angular resolution
(Lamb et al. 1998);
they do not suffer from central obscuration by the focal 
detector; they are lightweight and highly transparent. Moreover, the much more
isochronisity of the refractive system with respect to mirrors allows us
to take advantage of the short duration of the Cherenkov pulses ($\sim$ 2-5 ns)
lowering the signal integration time and reducing the night sky background,
with consequently decrease of the telescope energy threshold.
The optical 
performance required for GAW is obtained with a single flat Fresnel lens 
with a diameter of 3 meters, a thickness of 3 mm and a f\# of 1.4. The lens is made of 
ultraviolet transmitting acrylic with a transmittance of about 95\% 
from ultraviolet to near infrared. Chromatic 
aberration, present in the Fresnel lens, is minimized using a diffractive 
plane in the optics design.

\vspace{1cm} 
\centerline{\bf Tab. 1 - Baseline of the GAW design}
\begin{table}[h]
\hspace{3.3cm} 
\begin{tabular}{|l|c|}
\hline
Collector light        &   Fresnel lens   \\
Lens diameter          &   3 m            \\
Lens weight            &   30 kg          \\
Lens geometrical area  &   7.06 m$^2$    \\
Lens trasmittance      &   0.95 (300-600 nm) \\
Lens rms               &   0.1 degrees    \\
Focal length           &   4.2 m          \\
f\#                    &   1.4            \\
Focal plane detectors  &   MAPMT R7600-03-M64 \\
PMT number             &   300 \\
PMT working mode       &   single count mode \\
Pixel number           &   19200            \\
Focal plane pixel size &   4 arcmin \\
Total FoV              &   10.5 degrees      \\
Mount                  &   Alt-Alt           \\
\hline
\end{tabular}
\end{table}

The detector consists in an array of 300 multi-anode photomultipliers (MAPMT) 
manufactured by Hamamatsu, series R7600-03-M64 (Hamamatsu, 1999). Each 
photomultiplier is equipped with a bi-alkali photocatode and UV transmitting 
window that assure an average Quantum Efficiency of 20\% in 300-500 nm.  The 
MAPMT is organized as an array of 8$\times$8 
independent pixels.
The pixel size is $\sim$ 5 mm, 
corresponding to an angular size  
of $\sim$ 4 arcmin. The total array covers a FOV of 10.5 degrees. 
The MAPMT R7600-03-M64 has a geometric dead area of the order of 50\%; 
a suitable light collector system is therefore  
placed above the MAPMT array to lead photons directed to the dead space
to the sensitive area.
In GAW, thanks to the reduced pixel size, the MAPMTs 
can operate in single photoelectron count mode.
In such a working mode the noise and gain differences are negligible and it is 
possible to decrease the minimum number of photoelectrons necessary to apply 
the 
imaging background rejection technique. Small pixel size is actually required 
to minimize the probability of photoelectrons pile-up within intervals 
shorter than the dead time ( $\sim$ 5 ns). The imaging background rejection 
technique can be applied to images containing a minimum number of $\sim$40 
photoelectrons. This allows to lower the energy threshold to a value of 
$\sim$ 200 GeV, in spite of the relatively 
small dimension of the Cherenkov light collector.

\begin{figure}
\epsfysize=7cm 
\hspace{2.0cm}\epsfbox{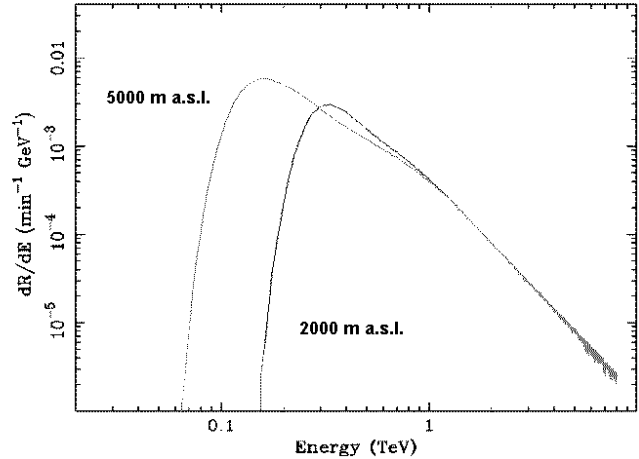} 
\caption[h]{Differential detection rate of the Crab Nebula expected with GAW 
located at two different observing level. The peak of the curves gives the low 
energy threshold of the telescope.}
\end{figure}

\section{GAW performance}

Fig. 2 shows the differential detection rate 
of the Crab Nebula as expected to be observed by GAW located at two possible 
different observing levels. We define the detection rate as the rate of 
gamma-ray events remaining after the cosmic-ray background rejection. The 
energy threshold of GAW, defined as the energy corresponding to the maximum 
of differential detection rate, 
will be about 150 and 320 GeV for the 2000 and 5000 meters a.s.l. observing 
level, respectively.

\begin{figure}
\epsfysize=7cm 
\hspace{2.0cm}\epsfbox{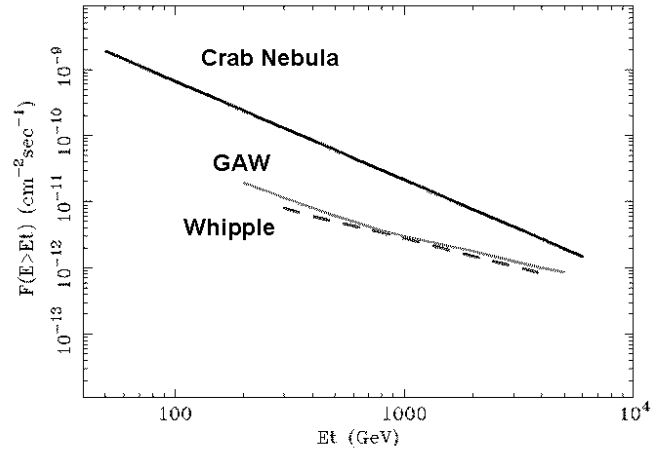} 
\caption[h]{The sensitivity (5$\sigma$) detection of GAW for point-like 
sources in 50 hours observations. In figure are also reported for comparison 
the point source sensitivity of Whipple (Weeks et al. 1989) and the integrated 
Crab Nebula flux.}
\end{figure}

The performance of GAW is summarized by its integrated flux sensitivity as 
function of the energy. Fig. 3 shows the minimum integrated flux for detection 
of a 5 $\sigma$ excess, with at least more than 10 photons, in 50 hours of 
observation of a gamma-ray source with a Crab-like spectrum (dN/dE $\propto$ $E^{-2.5}$).
For comparison, the Whipple sensitivity is shown in the figure together with the 
integrated Crab Nebula flux. GAW, thanks to its finely segmented 
focal plane detector operating in single count mode, has a low energy threshold 
and flux sensitivity comparable to the Whipple telescope (Weeks et al. 1989), in spite of the 
one order lower geometrical area of the Cherenkov light collector.

GAW will be the first wide FOV imaging Cherenkov telescope. It is suitable to 
survey interesting sky regions as the Galactic Center, the Galactic Plane, the
 Magellanic Clouds, etc.. Moreover, pointed observations will be performed with 
good sensitivity and with a
low energy threshold comparable to that of the next imaging Cherenkov 
telescopes.

\end{document}